\documentclass[nonacm,sigplan]{acmart}
\renewcommand\footnotetextcopyrightpermission[1]{}
\AtBeginDocument{
  }

\usepackage{xspace}
\usepackage{lstautogobble}

\setcopyright{acmlicensed}
\copyrightyear{2018}
\acmYear{2018}
\acmDOI{XXXXXXX.XXXXXXX}

\def \tool{\textsc{Kajal}\xspace}

\lstdefinestyle{JSXStyle}{
    basicstyle=\footnotesize\ttfamily,
    keywordstyle=\color{blue},
    rulecolor=\color{black},
    commentstyle=\color{gray},
    stringstyle=\color{darkgreen1},
    keywordstyle=\color{blue},
    frame=single,
    breaklines=true,
    breakatwhitespace=true,
    showstringspaces=false,
    tabsize=2,
    lineskip=1.2ex,
    float
}

\begin{document}

\title[]{\tool: Extracting Grammar of a Source Code Using Large Language Models}

\author{Mohammad Jalili Torkamani}
\email{MJaliliTorkamani2@huskers.unl.edu}
\orcid{0009-0002-3163-4850}
\affiliation{
  \institution{University of Nebraska-Lincoln}
  \city{Lincoln}
  \state{Nebraska}
  \country{USA}
}

\definecolor{darkgreen}{rgb}{0, 0.5, 0}
\definecolor{darkgreen1}{RGB}{37, 151, 19}
\definecolor{brown}{RGB}{139, 69, 19}
\definecolor{darkbrown}{rgb}{0.396, 0.263, 0.129}

\renewcommand{\shortauthors}{Mohammad Jalili Torkamani}

\begin{abstract}
Understanding and extracting the grammar of a domain-specific language (DSL) is crucial for various software engineering tasks; however, manually creating these grammars is time-intensive and error-prone. This paper presents \tool, a novel approach that automatically infers grammar from DSL code snippets by leveraging Large Language Models (LLMs) through prompt engineering and few-shot learning. \tool dynamically constructs input prompts, using contextual information to guide the LLM in generating the corresponding grammars, which are iteratively refined through a feedback-driven approach. Our experiments show that \tool achieves 60\% accuracy with few-shot learning and 45\% without it, demonstrating the significant impact of few-shot learning on the tool's effectiveness. This approach offers a promising solution for automating DSL grammar extraction, and future work will explore using smaller, open-source LLMs and testing on larger datasets to further validate \tool's performance.
\end{abstract}



\maketitle

\section{Introduction}
\label{sec:introduction}

Grammars have many applications in engineering and computer science \cite{841160}. They play an essential role in defining the structure that code must follow to be syntactically correct and interpretable by compilers \cite{waite2012compiler, grune2012modern} or interpreters \cite{mak2011writing}. Grammars provide developers with a standardized way to write and read code, ensuring reliability in compilation, execution, and maintenance \cite{lohmann2009language}. Moreover, they facilitate the creation of domain-specific languages (DSLs) \cite{10.1145/1118890.1118892, fowler2010domain}, enabling tailored solutions for specific tasks in software and enhancing both productivity and code quality. Understanding and defining the grammar of such languages is crucial for enabling advanced tooling, such as parsers, syntax highlighters, and transpilers.

However, extracting the grammar of an arbitrary DSL poses a significant challenge due to the complexity and diversity of these languages, which often deviate from the conventions of more familiar programming languages. For example, JSX, a DSL widely used with JavaScript, illustrates this complexity. Listing \ref{lst:jsx_sample} shows an example code snippet for this language. Although JSX shares some features with JavaScript, it allows developers to write HTML-like syntax directly within their JavaScript code. In contrast, standard JavaScript typically requires using the \texttt{document.createElement} method to create HTML elements (Listing \ref{lst:js_sample}). Consequently, the grammar of JSX differs significantly from that of JavaScript.

\begin{lstlisting}[style=JSXStyle,showstringspaces=false,tabsize=2,label=lst:jsx_sample,caption= Example of a JSX Code.,captionpos=b, float=h, basicstyle=\scriptsize]
@kimportk@ React @kfromk@ @s'react's@;

@kconstk@ MyComponent = () => {
  @kreturnk@ (
    @b<div>
      <h1>Hello, World!</h1>
      <p>This is a paragraph in JSX.</p>
    </div>b@
  );
};

@kexport defaultk@ MyComponent;

\end{lstlisting}

\begin{lstlisting}[style=JSXStyle,showstringspaces=false,tabsize=2,label=lst:js_sample,caption= Corresponding Javascript Code.,captionpos=b, float=h, basicstyle=\scriptsize]
@kimportk@ React @kfromk@ @s'react's@;

@kconstk@ MyComponent = () => {
  @kreturnk@ React.createElement(
    @s'div's@,
    @knullk@,
    React.createElement(@s'h1's@, @knullk@, @s'Hello, World!'s@),
    React.createElement(@s'p's@, @knullk@, @s'This is a paragraph in JSX.'s@)
  );
};

@kexport defaultk@ MyComponent;

\end{lstlisting}

In addition to their syntactical differences, the underlying rules governing these languages are not always well-documented or consistent, especially in legacy or ad-hoc DSLs used in specific industries \cite{mendez2016reverse}. Real-world scenarios, such as the intricate rules of a trading DSL for finance \cite{suzuki2023constructing} or a markup language customized for medical records \cite{vieru2024domain}, exemplify this complexity. As a result, manually extracting and formalizing grammar in such cases would be prohibitively time-consuming and error-prone.

To resolve this issue, several traditional approaches are employed for extracting the grammar of programming languages, including manual grammar specification, lexical and syntactic analysis \cite{lammel2001semi}, statistical \cite{jain2004interactive}, or machine learning-based grammar induction \cite{10.1145/3383458}, and grammar mining from existing code. While manual specification allows for precise control over grammar, it is labor-intensive and depends heavily on expert knowledge. Lexical and syntactic analysis can generate grammars from existing codebases but often results in incomplete or overly complex outputs. Statistical methods, while scalable, require large amounts of labeled data for supervised learning and may not yield grammars that are easily interpretable by humans. Grammar mining focuses on analyzing existing source code but may produce grammars tailored to specific codebases, lacking generalizability. Thus, a feasible approach to extracting the grammar of arbitrary DSLs would streamline the construction of these languages systematically, aiding in the design of parsers, transpilers, and static analysis tools. Such automated methods would also enhance the accessibility and integration of DSLs across various fields and domains.

Large language models (LLMs) \cite{chang2024survey} present a promising approach for grammar extraction and offer a flexible and adaptive method. LLMs can learn from a diverse range of programming examples and natural language descriptions, enabling them to infer complex grammar patterns. Additionally, they can be fine-tuned with minimal labeled data, facilitating efficient grammar extraction across various contexts while reducing the manual effort required.

In this paper, we introduce \tool as an innovative approach to extract the grammar of domain-specific languages. By employing prompt engineering techniques and providing relevant examples through few-shot learning, \tool assists LLMs in extracting grammar from relatively complex source code. Our approach is automated and does not require implementing LLMs locally; instead, it leverages their Application Programming Interface (APIs) for inference, resulting in a more powerful and cost-effective solution in terms of inference time and resource requirements. Furthermore, it eliminates the need for labeled data or training, making it more convenient than traditional machine learning-based approaches. To evaluate our method, we collected a dataset that the LLM has not encountered before. We divided this dataset into two distinct subsets: one for few-shot learning and the other for evaluation. We employed various quantitative and qualitative metrics to assess \tool's accuracy and will discuss its strengths and weaknesses throughout the paper.

Most state-of-the-art models have access to online resources. Therefore, we must ensure the LLM has not seen our collected evaluation dataset during its training. For this research, we used the GPT-3.5 \cite{ye2023comprehensive} model (hereby called GPT-3), which has a constant cut-off date. This choice allows us to select DSL code snippets as an evaluation set whose grammar has never been publicly released before the LLM's cut-off date. Consequently, we can be confident that the LLM has not been exposed to our dataset during its training.

The paper is organized as follows: Section \ref{sec:example} discusses a practical example of what problem our tool tries to resolve, how it works, and why its results are useful. Section \ref{sec:technical_details} will present the underlying technical details behind \tool, its implementation, and how it was developed, along with the limitations of the tool, and its requirements. In Section \ref{sec:evaluation} we present a discussion of the experiments, and interpret the results of our conducted experiments, along with some statistical details about the datasets. Section \ref{sec:related_works} discusses some of the conducted related works in the LLMs and grammar extraction area, categorized by the approach they follow. Finally, the paper concludes in Section \ref{sec:conclusion}, summarizing key findings, their significance, and potential future work.
\section{Example}
\label{sec:example}

In this section, we present a practical example to demonstrate the process of generating grammar for a DSL using \tool. This example provides an overview of \tool's purpose, outputs, and operational approach, as well as the benefits of our proposed method.

\tool is designed as an end-to-end tool that generates grammar for a given code snippet extracted from the dataset. This means that it takes two datasets as input and produces an output JSON file containing the inferred grammar for each dataset record, along with evaluation metrics that indicate performance and accuracy. The approach incorporates prompt engineering, dynamically generating a prompt for the LLM based on the input code snippet to effectively produce grammar. Additionally, it performs few-shot learning by providing three similar code snippets and their corresponding grammar to provide the LLM with some contextual insight.

A key component of \tool's operation is a feedback-driven iterative process along with few-shot prompting—a technique in prompt engineering that enhances inference accuracy through iterative prompt refinement. Recent research \cite{ye2022unreliability, meyerson2023language} has demonstrated the effectiveness of few-shot prompting in increasing inference accuracy. In \tool, this is implemented by creating a prompt from a code snippet, which the LLM uses to generate grammar in a specified output format. After the initial inference, the inferred grammar is tested by parsing the input code snippet. If parsing succeeds, the record is marked as a correct inferred grammar, and the process moves to the next dataset record. If parsing fails, \tool performs few-shot prompting. This means that it extracts the error message from the parsing library and feeds it back to the LLM as additional context for prompt correction. This feedback loop continues for up to ten iterations, providing the error message for each unsuccessful inference until a correct grammar (which can parse the given code snippet from the dataset) is generated or the attempt is marked as failed.

To illustrate this process, we consider a custom code snippet comprising a set of statements forming an arbitrary DSL where each statement starts with a capital letter \textit{K}. As this DSL is unique and unfamiliar to the LLM, the LLM has no prior exposure to its syntax or grammar. Therefore, the LLM must derive patterns from the code snippet itself rather than relying on its own training data.

\begin{lstlisting}[style=JSXStyle,showstringspaces=false,tabsize=2,label=lst:working_sample,caption= Example a DSL Code Snippet.,captionpos=b, float=h, basicstyle=\scriptsize]
Kset x = 5
Kadd x, 3
Kmul 10, 2
Kpow 2, 3
Kmod 10, 3
Ksqrt 16
Klog 100
Kfib 7
\end{lstlisting}

As shown in Listing \ref{lst:working_sample}, the code contains eight statements, each beginning with a capital \textit{K}, performing basic mathematical operations, assignments, and functions like Fibonacci. Some statements require two integer inputs, while others need only one. To process this code, \tool generates a prompt with instructions guiding the LLM to produce a grammar that can parse it. These instructions specify the input \tool provides, the expected output format, and the template for easier grammar extraction from the LLM's output. Additionally, three code snippets and grammar pairs are included for few-shot learning to assist the LLM's inference.

\begin{lstlisting}[style=JSXStyle,showstringspaces=false,tabsize=2,label=lst:working_sample_grammar,caption= Generated Grammar - First Iteration.,captionpos=b, float=h, basicstyle=\scriptsize]
    start: statement+

    statement: "Kadd" expr "," expr           -> add
             | "Ksub" expr "," expr           -> subtract
             | "Kmul" expr "," expr           -> multiply
             | "Kdiv" expr "," expr           -> divide
             | "Kmod" expr "," expr           -> mod
             | "Kpow" expr "," expr           -> power
             | "Ksqrt" expr                    -> sqrt
             | "Klog" expr                     -> log
             | "Kfac" expr                     -> factorial
             | "Kfib" expr                     -> fibonacci
             | "Kset" NAME "=" expr            -> set_var

    expr: NUMBER                            -> number
        | NAME                              -> variable
        | "(" expr ")"                     -> group
        | expr "+" expr                     -> add
        | expr "-" expr                     -> subtract
        | expr "*" expr                     -> multiply
        | expr "/" expr                     -> divide

    %import common.NUMBER                   // Import number token
    %import common.WS                       // Import whitespace
    @eCNAMEe@: /[a-zA-Z_][a-zA-Z0-9_]*/          // Define NAME token (variable names)
    %ignore WS                              // Ignore whitespace
\end{lstlisting}

According to Listing \ref{lst:working_sample_grammar}, the inferred grammar has 19 rules divided into three sections, each labeled with aliases and documented with comments. This initial output shows that \tool successfully identified DSL patterns, such as argument structures, whitespace rules, and necessary symbols. To validate the grammar, \tool uses the \textit{lark} library to parse the code snippet. After testing the inferred grammar, a minor issue is detected—a variable named \textit{CNAME} is incorrectly used as \textit{NAME} elsewhere, resulting in a parsing exception due to mismatched rules.

\begin{lstlisting}[style=JSXStyle,showstringspaces=false,tabsize=2,label=lst:working_sample_grammar_corrected,caption= Generated Grammar - Second Iteration.,captionpos=b, float=h, basicstyle=\scriptsize]
    start: statement+

    statement: "Kadd" expr "," expr           -> add
             | "Ksub" expr "," expr           -> subtract
             | "Kmul" expr "," expr           -> multiply
             | "Kdiv" expr "," expr           -> divide
             | "Kmod" expr "," expr           -> mod
             | "Kpow" expr "," expr           -> power
             | "Ksqrt" expr                    -> sqrt
             | "Klog" expr                     -> log
             | "Kfac" expr                     -> factorial
             | "Kfib" expr                     -> fibonacci
             | "Kset" NAME "=" expr            -> set_var

    expr: NUMBER                            -> number
        | NAME                              -> variable
        | "(" expr ")"                     -> group
        | expr "+" expr                     -> add
        | expr "-" expr                     -> subtract
        | expr "*" expr                     -> multiply
        | expr "/" expr                     -> divide

    %import common.NUMBER                   // Import number token
    %import common.WS                       // Import whitespace
    NAME: /[a-zA-Z_][a-zA-Z0-9_]*/          // Define NAME token (variable names)
    %ignore WS                              // Ignore whitespace
\end{lstlisting}

To address this, \tool performs another iteration (few-shot prompting), using the error message as a new context for the LLM and requesting a revised inference. The corrected output, shown in Listing \ref{lst:working_sample_grammar_corrected}, consistently defines the variable as \textit{NAME}, yielding a correct grammar. This revised grammar successfully parses the input code snippet, confirming alignment with its structure.

This example highlights \tool's capability to generate accurate, usable grammars for arbitrary DSLs through prompt engineering, few-shot learning, few-shot prompting, and an iterative feedback-driven approach. Ultimately, this method showcases an automatic process that requires no human intervention to identify patterns in a given code snippet and generate compatible grammar.
\section{Technical Details}
\label{sec:technical_details}

In this section, we provide a comprehensive description of \tool's implementation, operational workflow, and the types of information it processes. Additionally, we include a schematic diagram to visually represent the workflow, aiding in a clearer understanding of \tool's functionalities.

\begin{figure*}
    \centering
    \includegraphics[width=0.95\textwidth]{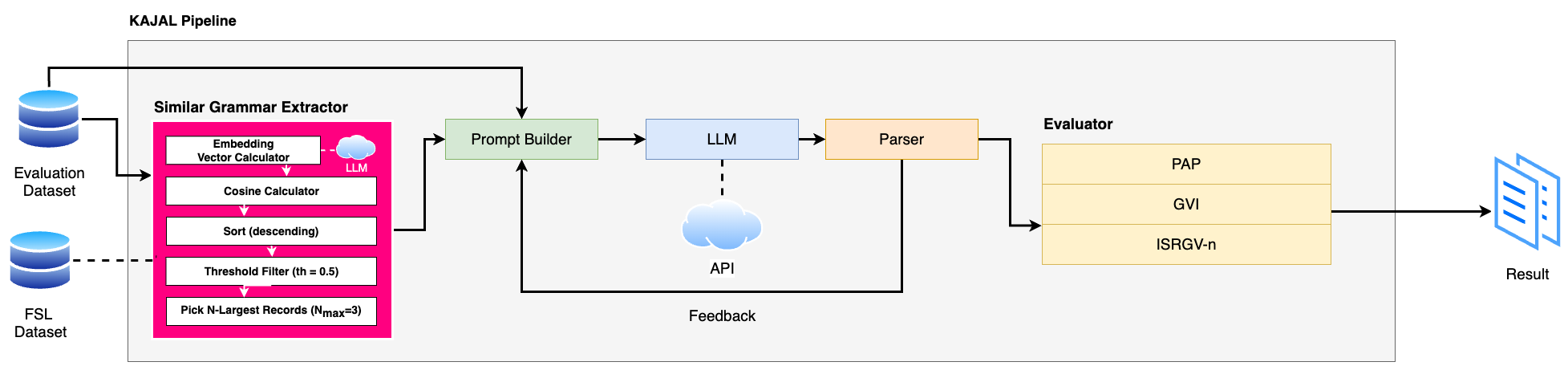}
    \caption{\textbf{\tool Workflow.}}
    \label{fig:pipeline_diagram}
\end{figure*}

Overall, \tool is an end-to-end tool pipeline implemented in Python, which can be executed using the same language. It requires two input datasets: one for evaluation and another for few-shot learning. The end-to-end feature ensures that the tool processes the inputs and produces outputs automatically. The first input dataset, the evaluation dataset, is used to assess the performance of \tool. The second input dataset, the few-shot learning dataset (FSL dataset), is used to extract similar records. This process will be elaborated later. Upon execution, the pipeline generates an output JSON file containing the inferred grammar for the input dataset's records, along with quantitative results computed for the overall inference. Moreover, \tool employs a feedback-driven iterative approach, which means that it iteratively refines its inference by providing the LLM with the latest error messages from the parser as feedback when parsing the input record with the inferred grammar.

The pipeline comprises five sequential components. Each component processes its inputs and passes the results to the next component. Figure \ref{fig:pipeline_diagram} illustrates the \tool workflow.

As shown in Figure \ref{fig:pipeline_diagram}, the pipeline iterates over the evaluation dataset. For each record, the following process occurs:

The DSL code snippet is extracted from the evaluation dataset.
This snippet is passed to the \textit{Similar Grammar Extractor} component, which identifies the three most similar code snippets from the FSL dataset along with their associated grammars. To compute similarity during prompt engineering, \tool vectorizes both the candidate snippet and each snippet in the FSL dataset, padding them to ensure uniform sizes. It then applies the cosine similarity formula, sorts the results in descending order, and uses a minimum similarity threshold of 0.5:
\[
\text{{cosine similarity}}(\mathbf{A}, \mathbf{B}) = \frac{\mathbf{A} \cdot \mathbf{B}}{\|\mathbf{A}\| \times \|\mathbf{B}\|}
\]

This threshold balances between identical records, and dissimilar records, ensuring a fair selection of snippets.

\begin{figure}
    \includegraphics[width=0.45\textwidth]{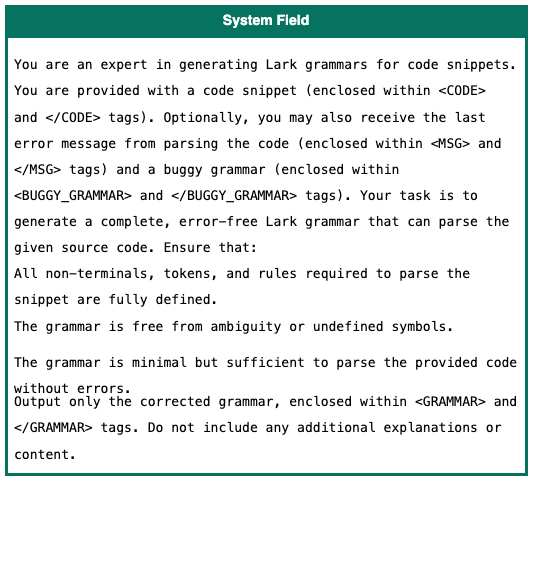}
    \caption{\textbf{System Field.}}
    \label{fig:system_field}
\end{figure}

The extracted similar code snippets, along with the original input snippet, are then passed to the \textit{Prompt Builder} component. This component constructs a prompt for the next component, the LLM. To construct the prompt, the system field (illustrated in Figure \ref{fig:system_field}) is used. This field contains instructions guiding the LLM in understanding the input and generating the required output in the desired format. For the user field, the source snippet and, optionally, the feedback message (along with the most recent invalid grammar) are included. In few-shot learning experiments, similar code snippets and their grammar are passed to the LLM as user-assistant pairs, enabling the model to learn from limited examples without requiring additional training data.

After constructing the LLM fields (system, user, and assistant), the input is passed to the \textit{LLM} component. This component uses the GPT-3 model (with default configurations) to perform inference. The GPT-3 model is selected due to its fixed knowledge cut-off date, ensuring that it does not utilize information about grammar released after this date. This is critical for evaluating \tool's ability to infer grammar from novel and unseen code snippets. Therefore, the use of GPT-4 or GPT-4o models is avoided to preserve the research's focus. The \textit{LLM} component utilizes APIs for inference, ensuring accessibility and scalability without requiring heavy computational resources.

\begin{figure}[b]
    \includegraphics[width=0.45\textwidth]{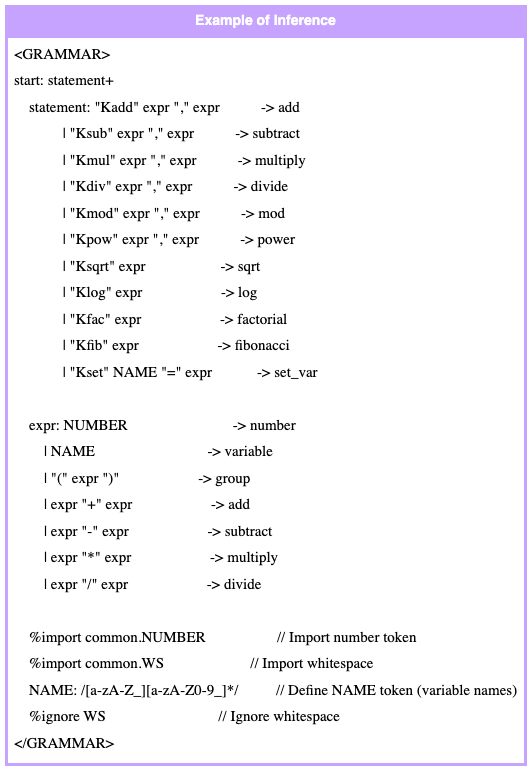}
    \caption{\textbf{Example of Inference.}}
    \label{fig:inference_example}
\end{figure}

The inferred output is then processed by the \textit{Parser} component, which extracts the grammar from the LLM’s response by identifying content enclosed within <GRAMMAR> and </GRAMMAR> tags using regular expressions. These tags improve output accuracy and simplify grammar extraction. Figure \ref{fig:inference_example} illustrates an example of inference made by \tool. The parser also tests the grammar by attempting to parse the input code snippet using the \textit{Lark} library with an \textit{lalr} parser. Then, each inference result is evaluated by the \textit{Evaluator} component based on predefined metrics (detailed in Section \ref{sec:evaluation}). All results are saved as a single comprehensive JSON file for further analysis.

One key feature of \tool is its feedback-driven iterative approach, designed to enhance inference accuracy. If the parser encounters errors, the inferred buggy grammar, along with the error message generated by the \textit{Lark} library, is returned to the LLM for refinement. This feedback serves as additional context for the LLM. The process is repeated for up to ten iterations. Regardless of success or failure, all details are logged in the output JSON file.

Despite the advantages of our work over related works (described in Section \ref{sec:related_works}), \tool has limitations. First, it does not ensure semantic accuracy in inferred grammars, potentially leading to ambiguities. This limitation arises because grammar extraction often requires domain-specific knowledge. Second, \tool lacks mechanisms to validate the quality of feedback or mitigate biases in its iterative approach. Finally, for large codebases requiring extensive token usage, \tool may face limitations depending on the adopted LLM's capabilities and constraints.

\section{Evaluation}
\label{sec:evaluation}

The purpose of this section is to demonstrate how \tool performs when applied to our evaluation dataset. We aim to showcase our evaluation methodology, establish clear evaluation goals, and introduce the metrics that allow us to assess \tool's efficiency through qualitative and quantitative measures.

\subsection{Dataset Collection}
To evaluate \tool's effectiveness, the selection of an appropriate dataset is essential. As discussed in Section \ref{sec:introduction}, we use the GPT-3 model to maintain consistency, ensuring its cut-off date does not shift over time. For a fair evaluation, a selected dataset ideally should not be seen by the LLM during training, ensuring that \tool's grammar inference is unbiased and based solely on the input code. We also collected an additional dataset (independent of release dates), to support few-shot learning. This secondary dataset is excluded from the evaluation dataset but provides context for the LLM.

To the best of our knowledge, by the time we wrote this paper, we could not find any appropriate datasets released after the cut-off date. Therefore, we used the ChatGPT-4o model to create 23 code snippets. Ten of these code snippets are similar to well-known programming languages but contain slight differences in syntax. The remaining snippets cover various domains, including networking, data transformation, user permissions, task management, mathematical expressions, and more. These code snippets are stored in a single JSON file. We then split this dataset into two distinct subsets: an evaluation set (containing 20 samples) with release dates after the LLM's cut-off, and a few-shot learning dataset (containing 3 samples) with no cut-off restrictions. By connecting these datasets to our end-to-end tool, \tool can sequentially process each record in the evaluation set, perform inference, evaluate the results, and store them in a comprehensive JSON file.

\subsection{Methodology}
We conduct the experiments in two distinct variants. The first is our baseline, where we run \tool without any few-shot learning, relying solely on the LLM’s default inference capability using prompt engineering. This approach reveals the base performance of our chosen LLM. We will compare the results of the second variant with this baseline to gauge improvement.

In the second variant, we adopt a few-shot learning approach, where similar code snippets along with their grammar are fed to the LLM, enabling the LLM to have more accurate inference and generate a valid grammar that can parse the given source code.

\subsection{Evaluation Metrics}

To assess \tool's accuracy and performance, we defined a set of metrics that help us gauge the effectiveness of our approach and tool. These metrics also assist in identifying areas for improvement. The metrics are as follows:

\begin{itemize}
    \item \textbf{Parsing Accuracy Percentage (PAP)}: This metric reflects the percentage of code snippets for which \tool inferred the correct grammar. It indicates the accuracy and efficiency of our approach in generating grammars that can parse the given code snippet. PAP is calculated as follows:
    \[
    \text{PAP} = \frac{\text{Number of Correctly Parsed Code Snippets}}{\text{Total Number of Code Snippets}} \times 100
    \]

    \item \textbf{Grammar Validity Index (GVI)}: This metric measures whether the inferred grammar (and generally, output) is structurally valid and can be processed by a specific parsing tool (e.g., \textit{lark}), regardless of the code snippet. It evaluates the structural correctness of the generated grammar, helping us understand the LLM’s capabilities in generating grammars compatible with the parsing library. GVI is calculated as follows:
    \[
    \text{GVI} = \frac{\text{Number of Structurally Valid Grammars}}{\text{Total Number of Code Snippets}} \times 100
    \]

    \item \textbf{Iteration Success Rate of Grammar Correctness in the n\textsuperscript{th} Iteration (ISRGC-n)}: This measures the success rate of generating correct grammar by the n\textsuperscript{th} iteration. It helps us understand the effectiveness of the feedback-driven approach in producing correct grammar.
\end{itemize}

We will also conduct manual evaluations to assess the qualitative aspects of the generated grammar. This analysis will help categorize any failures and identify areas for improvement, guiding future enhancements.

\subsection{Results and Analysis}
After conducting the experiments, the following statistics have been achieved:

\begin{table*}[ht]
\centering
\begin{tabular}{|l|c|c|}
\hline
\textbf{Metric} & \textbf{Without Few-Shot Learning} & \textbf{With Few-Shot Learning} \\
\hline
Total Number of Items & 20 & 20 \\
Valid Inferences & 9 & 12 \\
Correct Inferences & 9 & 12 \\
Invalid Inferences & 11 & 8 \\
Incorrect Inferences & 11 & 8 \\
GIV (\%) & 45.0 & 60.0 \\
PAP (\%) & 45.0 & 60.0 \\
\hline
\end{tabular}
\caption{Comparison of \tool performance with/without few-shot learning.}
\label{table_metrics}
\end{table*}

\begin{figure}[b]
    \includegraphics[width=0.45\textwidth]{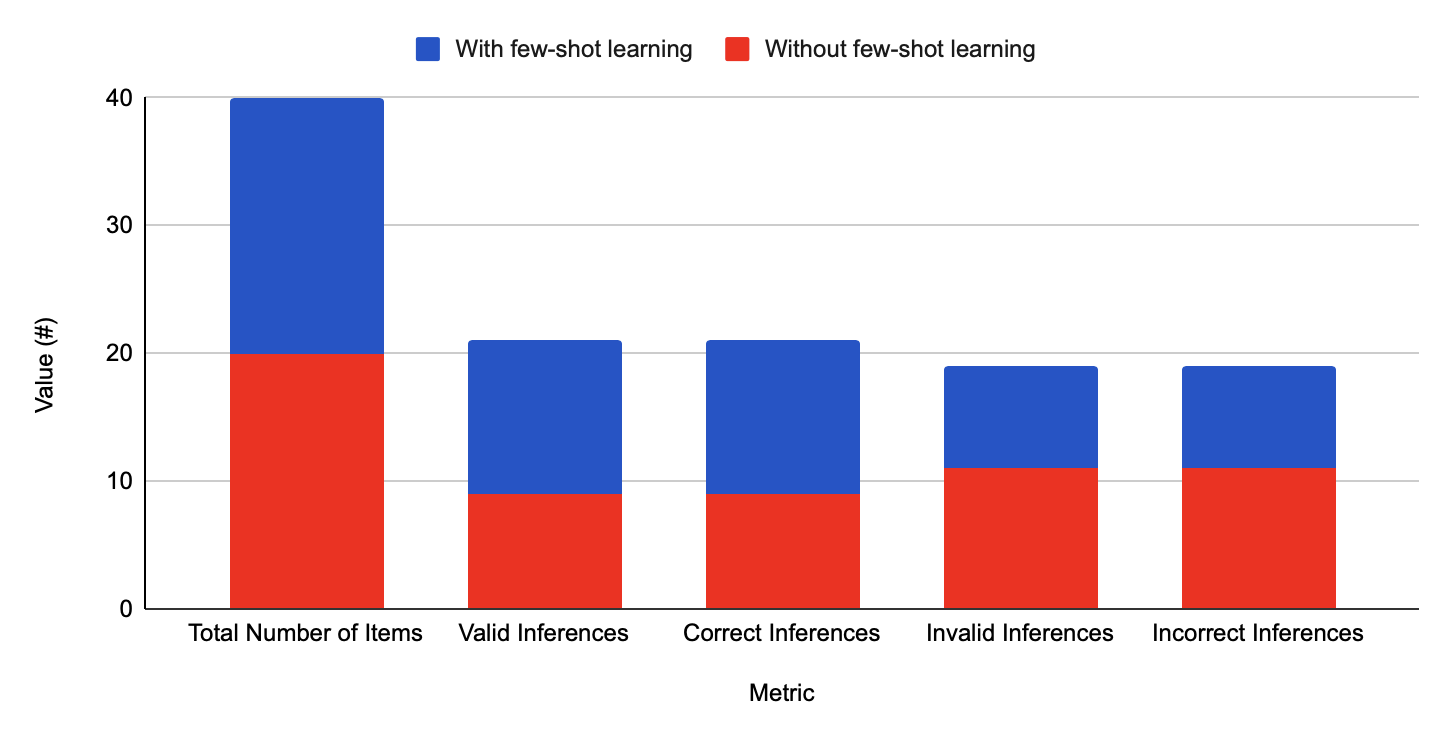}
    \caption{\textbf{\tool Experiments' Results.}}
    \label{fig:metrics}
\end{figure}

\begin{figure}[b]
    \includegraphics[width=0.48\textwidth]{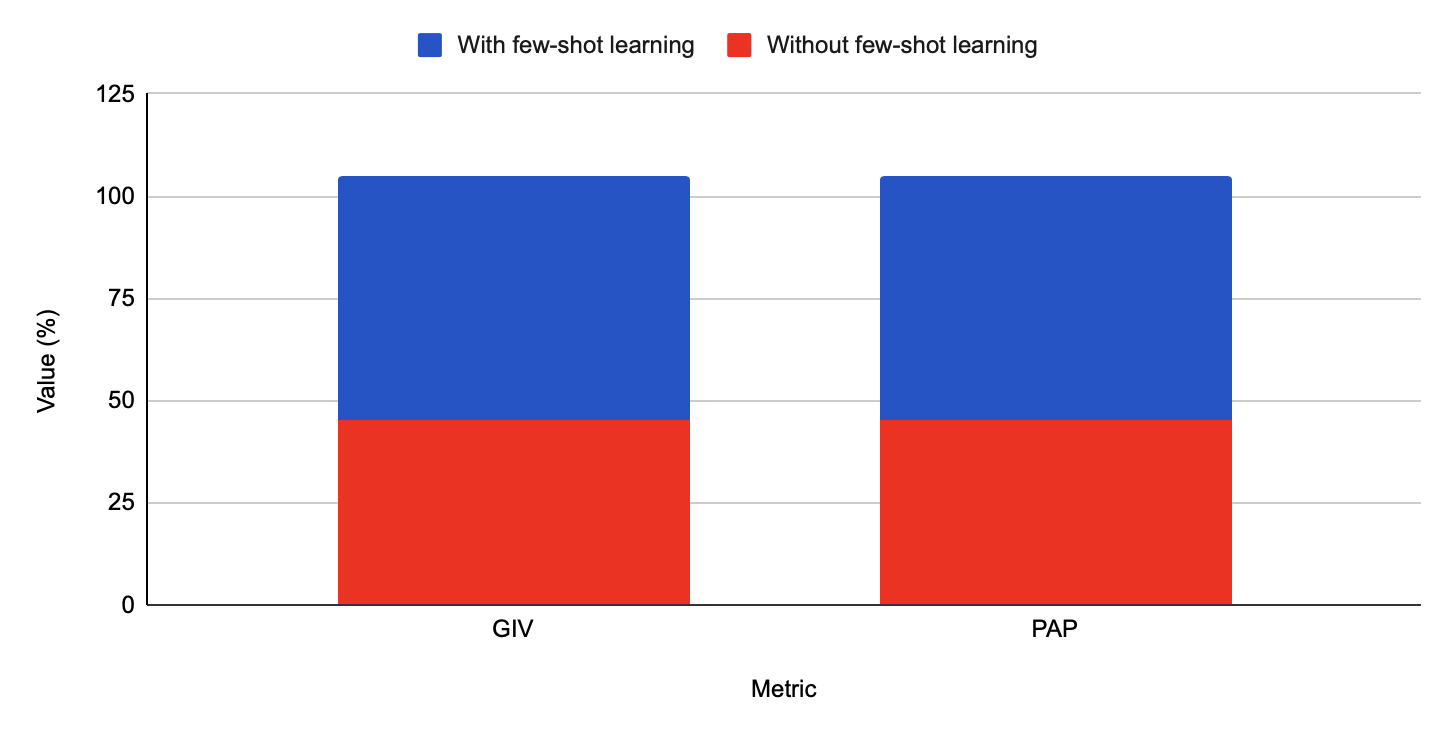}
    \caption{\textbf{\tool GIV and PAP Metrics.}}
    \label{fig:percentage_metrics}
\end{figure}

The results presented in Table \ref{table_metrics} and Figures \ref{fig:metrics} and \ref{fig:percentage_metrics} highlight the impact of incorporating few-shot learning into the grammar extraction process for a DSL using LLMs. Without few-shot learning, the system achieved a validity accuracy of 45.0\% and a correctness accuracy of 45.0\%, reflecting a relatively low rate of valid and correct inferences. However, when few-shot learning was introduced, both validity and correctness accuracy improved significantly to 60.0\%. This increase suggests that few-shot learning helps the model generalize better from limited examples, leading to more accurate inferences and a reduction in the number of invalid or incorrect outputs.

These findings underscore the value of few-shot learning in enhancing the performance of \tool for grammar extraction. The improvement in both valid and correct inferences (from 9 to 12) coupled with the reduction in invalid and incorrect inferences (from 11 to 8) indicates that few-shot learning contributes to a more effective and reliable output. This suggests that grammar extraction from DSLs, leveraging few-shot learning not only boosts accuracy but also reduces errors, making our approach a promising technique for grammar extraction from DSLs.

\begin{figure}[h]
    \includegraphics[width=0.45\textwidth]{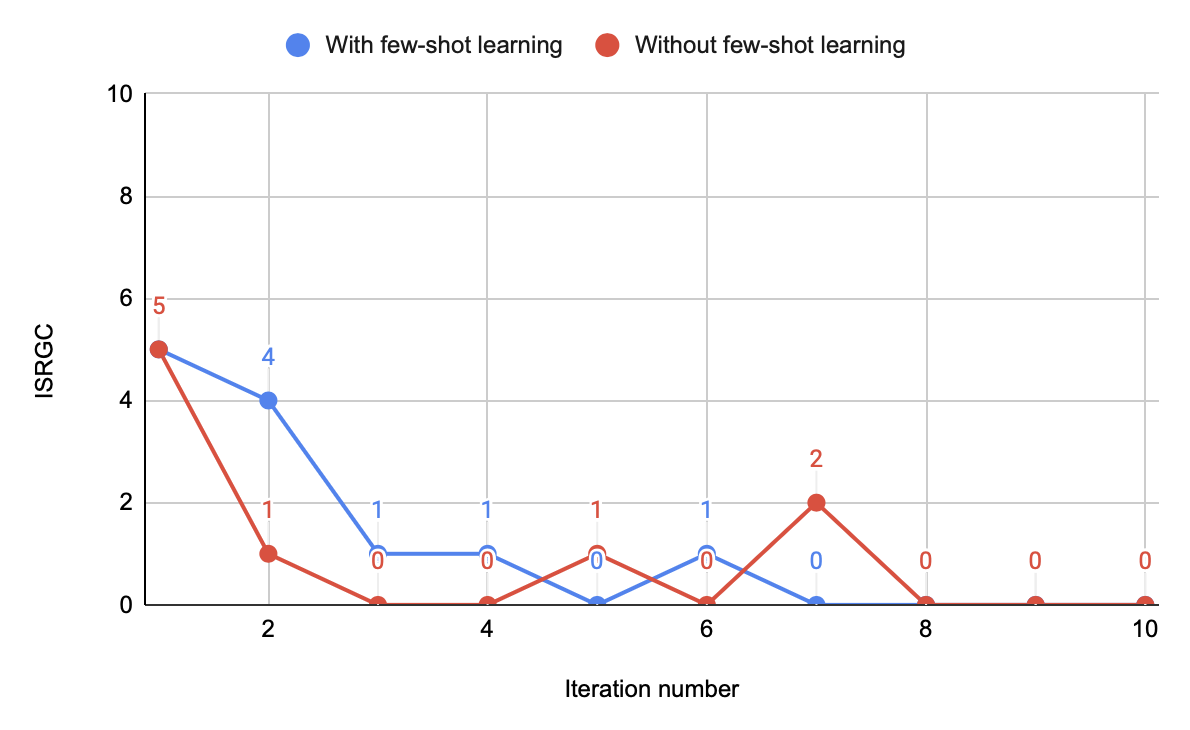}
    \caption{\textbf{Iteartion Success Rate.}}
    \label{fig:ISRGC}
\end{figure}

In terms of the iteration success rate of grammar correctness in the nth iteration (ISRGC-n), the following results have been computed (as depicted in Figure \ref{fig:ISRGC}):

\subsubsection{\textbf{ISRGC / With Few-Shot Learning}}:
The iterative approach with few-shot learning demonstrates a clear trend of generating correct inferences in the early iterations. In the first two iterations, the model produces 5 and 4 correct grammars, respectively, showing that it is able to refine its grammar generation effectively. However, as iterations continue, the number of correct inferences declines, converging to 0 by iteration 5. Despite this decline, the model still generates some correct inferences in later iterations (1 correct inference in iteration 6), which suggests that the iterative feedback approach helps the model gradually refine the grammar over time. In other words, the iterative process still provides opportunities for it to generate correct grammar in the next iterations.

\subsubsection{\textbf{ISRGC / Without Few-Shot Learning}}:
Without few-shot learning, the model shows a slightly different behavior. The first iteration produces 5 correct inferences, but there is a sharp drop to 1 correct inference in the second iteration, and no correct inferences are generated in subsequent iterations, with the exception of iterations 5 and 7, where 1 and 2 correct inferences appear respectively. Although this fluctuation shows the possibility of inferring correct grammars in later iterations, having low PAP shows the impact of few-shot learning in the \tool's total accuracy.

As a comparison of the described variants, the iterative approach proves to be essential in both scenarios, but especially in the case without few-shot learning. While the model’s performance declines over time in both cases, the iterative process allows it to continue making corrections and refining its grammar, ensuring that some correct inferences are generated even in later iterations. Without this iterative feedback loop, the model would not have been able to make further inferences for each record, hindering its ability to improve or adjust its grammar generation. Therefore, the iterative approach plays a crucial role in increasing the \tool's accuracy over time, particularly when few-shot learning is not present. 

The experiments for the attached datasets took less than 2 minutes to complete for each variant.

As a manual analysis of the experiment results, and according to our released artifacts, our analysis shows that the most frequent errors across all variants when making inferences are as follows:

\begin{itemize}
    \item Rule [RULE] used but not defined.
    \item Unexpected token [TOKEN].
    \item No terminal matches [VARIABLE] in the current parser context.
\end{itemize}

The error messages observed in the experiment results reveal critical insights into the challenges faced when inferring grammars with Lark. The most frequent issue, "Rule [RULE] used but not defined," indicates that the inferred grammar references rules that have not been declared, suggesting incomplete or inconsistent rule generation during inference. The error "Unexpected token [TOKEN]" points to scenarios where the parser encounters input sequences that deviate from the expected syntax, often due to incorrect token definitions or misplaced rules in the grammar. Finally, "No terminal matches [VARIABLE] in the current parser context" highlights mismatches between the input and the defined grammar's terminal symbols, signaling either gaps in coverage or conflicts in the tokenization process. Together, these errors underline the complexity of achieving fully functional and syntactically sound grammar, emphasizing the need for more robust inference strategies and systematic validation of generated rules.
\section{Related Works}
\label{sec:related_works}

Grammar induction has been an active area of research, with various approaches proposed to infer the structure of programming languages or domain-specific languages. These methods range from traditional algorithms to modern machine learning techniques, each addressing specific challenges such as balancing precision and generality or handling complex syntax. In this section, we discuss some of the most relevant research works that have explored grammar inference through computational methods, highlighting their contributions and limitations in comparison to our approach, which leverages Large Language Models (LLMs) to extract grammars from source code.

Jain et al. proposed an incremental approach \cite{interactive_grammar_extraction} for extracting the grammar from a source code (or the dialects) written in C, C++, and COBOL in an interactive way. Their approach starts with an approximate grammar and iteratively refines it by parsing source programs, identifying constructs not covered by the current grammar, and extending the grammar with rules sourced from a knowledge base or manually provided by the user. This work differs from our research in several ways. While their method relies on user interaction and a pre-defined knowledge base for refining the grammar, our research focuses on leveraging LLMs to extract grammar automatically. This shift eliminates the need for extensive manual intervention or pre-existing knowledge bases, enabling a more scalable and automated process. Additionally, \tool addresses not only dialects of existing languages but also explores the potential of LLMs to infer new syntactic patterns from examples, extending the scope beyond legacy systems to novel grammar extraction challenges.

Saha et al. proposed a programming language grammar inference system called Gramin \cite{Saha2011GraminAS}, which adapts grammar inference techniques traditionally used in the Natural Language Processing (NLP) domain to the programming language domain. Their approach includes an algorithm named Focus, designed to narrow the search space to non-terminals associated with parsing errors. Unlike \tool, which uses a feedback-driven iterative approach to refine the inferred rules, Gramin incrementally modifies the candidate grammar by adding new rules to resolve parsing errors. Gramin's methodology relies heavily on heuristics to prioritize rule selection and utilizes Prolog’s natural backtracking for efficient search and ranking of possible grammar rules. This makes Gramin less scalable compared to \tool's approach, where the use of LLMs can facilitate broader grammar extraction with potentially higher adaptability and automation.

In another research conducted by Lämmel et al. \cite{semi_automatic_recovery}, a semi-automated approach was proposed for extracting grammars of programming languages from existing artifacts such as compilers and language reference manuals. This method is particularly valuable for facilitating software renovation efforts, enabling the processing of legacy code where detailed grammar specifications might not be readily available. Furthermore, as most programming languages evolve, this approach supports the incremental extension of grammar definitions. The methodology involves multiple steps, starting with raw grammar extraction, followed by static error resolution, such as fixing unconnected non-terminals, and test-driven correction, where sample code is parsed to identify and address issues in the extracted grammar. Additional steps include modularization, beautification, and adaptation for specific use cases such as software renovation. A key distinction from \tool is that Lämmel et al.'s approach relies on pre-existing structured artifacts and adapts them for renovation and analysis. In contrast, \tool leverages LLMs to directly extract grammar from source code, potentially streamlining grammar acquisition in scenarios where artifacts are incomplete or unavailable.

The latest related work presented within this paper is a research conducted by Mernik et al., exploring the challenge of grammar induction using genetic programming for Domain-Specific Languages. The authors present a novel evolutionary algorithm that employs grammar-specific genetic operators for tasks like crossover and mutation, achieving grammar inference. The approach is validated through experiments, demonstrating its feasibility in generating parsers for DSLs. While the paper acknowledges the difficulty of inferring grammars for general-purpose programming languages, it successfully applies its method to DSLs, addressing syntax generation and, to some extent, semantics via attribute grammars. This work differs significantly from \tool, which leverages LLMs for extracting grammars from source code. Unlike the evolutionary algorithm approach used in this paper, which relies on structured genetic programming and predefined fitness criteria, \tool utilizes LLMs' contextual understanding and pattern recognition capabilities. Thus, our research enables a broader application scope, including complex and general-purpose languages, without requiring extensive manually designed evaluation functions.
\section{Conclusion}
\label{sec:conclusion}

In this paper, we introduce \tool as a novel approach for extracting the grammar of domain-specific languages. Unlike prior methods, particularly those based on machine learning, our approach leverages large language models as grammar generators through APIs, making it independent of their underlying implementations. Our experiments show the effectiveness of our approach, with 60\% accuracy when using few-shot learning and 45\% accuracy without it, highlighting the impact of few-shot learning on increasing the accuracy of \tool. In the future, we plan to use smaller, open-source LLMs and test our tool on larger datasets to further understand the effectiveness of our approach.
\section{Data Availability}
\label{sec:data_availability}

All artifacts including the source codes and experiments' results are available online \footnote{ \url{https://github.com/mohammadJaliliTorkamani/Kajal}}.

\bibliographystyle{acm_reference_format}
\bibliography{references}


\begin{thebibliography}{20}


\ifx \showCODEN    \undefined \def \showCODEN     #1{\unskip}     \fi
\ifx \showDOI      \undefined \def \showDOI       #1{#1}\fi
\ifx \showISBNx    \undefined \def \showISBNx     #1{\unskip}     \fi
\ifx \showISBNxiii \undefined \def \showISBNxiii  #1{\unskip}     \fi
\ifx \showISSN     \undefined \def \showISSN      #1{\unskip}     \fi
\ifx \showLCCN     \undefined \def \showLCCN      #1{\unskip}     \fi
\ifx \shownote     \undefined \def \shownote      #1{#1}          \fi
\ifx \showarticletitle \undefined \def \showarticletitle #1{#1}   \fi
\ifx \showURL      \undefined \def \showURL       {\relax}        \fi
\providecommand\bibfield[2]{#2}
\providecommand\bibinfo[2]{#2}
\providecommand\natexlab[1]{#1}
\providecommand\showeprint[2][]{arXiv:#2}

\bibitem[Chang et~al\mbox{.}(2024)]%
        {chang2024survey}
\bibfield{author}{\bibinfo{person}{Yupeng Chang}, \bibinfo{person}{Xu Wang}, \bibinfo{person}{Jindong Wang}, \bibinfo{person}{Yuan Wu}, \bibinfo{person}{Linyi Yang}, \bibinfo{person}{Kaijie Zhu}, \bibinfo{person}{Hao Chen}, \bibinfo{person}{Xiaoyuan Yi}, \bibinfo{person}{Cunxiang Wang}, \bibinfo{person}{Yidong Wang}, {et~al\mbox{.}}} \bibinfo{year}{2024}\natexlab{}.
\newblock \showarticletitle{A survey on evaluation of large language models}.
\newblock \bibinfo{journal}{\emph{ACM Transactions on Intelligent Systems and Technology}} \bibinfo{volume}{15}, \bibinfo{number}{3} (\bibinfo{year}{2024}), \bibinfo{pages}{1--45}.
\newblock


\bibitem[Fowler(2010)]%
        {fowler2010domain}
\bibfield{author}{\bibinfo{person}{Martin Fowler}.} \bibinfo{year}{2010}\natexlab{}.
\newblock \bibinfo{booktitle}{\emph{Domain-specific languages}}.
\newblock \bibinfo{publisher}{Pearson Education}.
\newblock


\bibitem[Grune et~al\mbox{.}(2012)]%
        {grune2012modern}
\bibfield{author}{\bibinfo{person}{Dick Grune}, \bibinfo{person}{Kees Van~Reeuwijk}, \bibinfo{person}{Henri~E Bal}, \bibinfo{person}{Ceriel~JH Jacobs}, {and} \bibinfo{person}{Koen Langendoen}.} \bibinfo{year}{2012}\natexlab{}.
\newblock \bibinfo{booktitle}{\emph{Modern compiler design}}.
\newblock \bibinfo{publisher}{Springer Science \& Business Media}.
\newblock


\bibitem[Jain et~al\mbox{.}(2004a)]%
        {jain2004interactive}
\bibfield{author}{\bibinfo{person}{Rahul Jain}, \bibinfo{person}{Sanjeev~Kumar Aggarwal}, \bibinfo{person}{Pankaj Jalote}, {and} \bibinfo{person}{Shiladitya Biswas}.} \bibinfo{year}{2004}\natexlab{a}.
\newblock \showarticletitle{An interactive method for extracting grammar from programs}.
\newblock \bibinfo{journal}{\emph{Software: Practice and Experience}} \bibinfo{volume}{34}, \bibinfo{number}{5} (\bibinfo{year}{2004}), \bibinfo{pages}{433--447}.
\newblock


\bibitem[Jain et~al\mbox{.}(2004b)]%
        {interactive_grammar_extraction}
\bibfield{author}{\bibinfo{person}{Rahul Jain}, \bibinfo{person}{Sanjeev~Kumar Aggarwal}, \bibinfo{person}{Pankaj Jalote}, {and} \bibinfo{person}{Shiladitya Biswas}.} \bibinfo{year}{2004}\natexlab{b}.
\newblock \showarticletitle{An interactive method for extracting grammar from programs}.
\newblock \bibinfo{journal}{\emph{Software: Practice and Experience}} \bibinfo{volume}{34}, \bibinfo{number}{5} (\bibinfo{year}{2004}), \bibinfo{pages}{433--447}.
\newblock
\urldef\tempurl%
\url{https://doi.org/10.1002/spe.568}
\showDOI{\tempurl}
\showeprint{https://onlinelibrary.wiley.com/doi/pdf/10.1002/spe.568}


\bibitem[Kieffer and Yang(2000)]%
        {841160}
\bibfield{author}{\bibinfo{person}{J.C. Kieffer} {and} \bibinfo{person}{En-Hui Yang}.} \bibinfo{year}{2000}\natexlab{}.
\newblock \showarticletitle{Grammar-based codes: a new class of universal lossless source codes}.
\newblock \bibinfo{journal}{\emph{IEEE Transactions on Information Theory}} \bibinfo{volume}{46}, \bibinfo{number}{3} (\bibinfo{year}{2000}), \bibinfo{pages}{737--754}.
\newblock
\urldef\tempurl%
\url{https://doi.org/10.1109/18.841160}
\showDOI{\tempurl}


\bibitem[L{\"a}mmel and Verhoef(2001)]%
        {lammel2001semi}
\bibfield{author}{\bibinfo{person}{Ralf L{\"a}mmel} {and} \bibinfo{person}{Chris Verhoef}.} \bibinfo{year}{2001}\natexlab{}.
\newblock \showarticletitle{Semi-automatic grammar recovery}.
\newblock \bibinfo{journal}{\emph{Software: Practice and Experience}} \bibinfo{volume}{31}, \bibinfo{number}{15} (\bibinfo{year}{2001}), \bibinfo{pages}{1395--1438}.
\newblock


\bibitem[Le et~al\mbox{.}(2020)]%
        {10.1145/3383458}
\bibfield{author}{\bibinfo{person}{Triet H.~M. Le}, \bibinfo{person}{Hao Chen}, {and} \bibinfo{person}{Muhammad~Ali Babar}.} \bibinfo{year}{2020}\natexlab{}.
\newblock \showarticletitle{Deep Learning for Source Code Modeling and Generation: Models, Applications, and Challenges}.
\newblock \bibinfo{journal}{\emph{ACM Comput. Surv.}} \bibinfo{volume}{53}, \bibinfo{number}{3}, Article \bibinfo{articleno}{62} (\bibinfo{date}{June} \bibinfo{year}{2020}), \bibinfo{numpages}{38}~pages.
\newblock
\showISSN{0360-0300}
\urldef\tempurl%
\url{https://doi.org/10.1145/3383458}
\showDOI{\tempurl}


\bibitem[Lohmann(2009)]%
        {lohmann2009language}
\bibfield{author}{\bibinfo{person}{Wolfgang Lohmann}.} \bibinfo{year}{2009}\natexlab{}.
\newblock \emph{\bibinfo{title}{On language processors and software maintenance}}.
\newblock \bibinfo{thesistype}{Ph.\,D. Dissertation}. \bibinfo{school}{Citeseer}.
\newblock


\bibitem[Lämmel and Verhoef(2001)]%
        {semi_automatic_recovery}
\bibfield{author}{\bibinfo{person}{R. Lämmel} {and} \bibinfo{person}{C. Verhoef}.} \bibinfo{year}{2001}\natexlab{}.
\newblock \showarticletitle{Semi-automatic grammar recovery}.
\newblock \bibinfo{journal}{\emph{Software: Practice and Experience}} \bibinfo{volume}{31}, \bibinfo{number}{15} (\bibinfo{year}{2001}), \bibinfo{pages}{1395--1438}.
\newblock
\urldef\tempurl%
\url{https://doi.org/10.1002/spe.423}
\showDOI{\tempurl}
\showeprint{https://onlinelibrary.wiley.com/doi/pdf/10.1002/spe.423}


\bibitem[Mak(2011)]%
        {mak2011writing}
\bibfield{author}{\bibinfo{person}{Ronald Mak}.} \bibinfo{year}{2011}\natexlab{}.
\newblock \bibinfo{booktitle}{\emph{Writing compilers and interpreters: a software engineering approach}}.
\newblock \bibinfo{publisher}{John Wiley \& Sons}.
\newblock


\bibitem[M{\'e}ndez-Acu{\~n}a et~al\mbox{.}(2016)]%
        {mendez2016reverse}
\bibfield{author}{\bibinfo{person}{David M{\'e}ndez-Acu{\~n}a}, \bibinfo{person}{Jos{\'e}~A Galindo}, \bibinfo{person}{Benoit Combemale}, \bibinfo{person}{Arnaud Blouin}, \bibinfo{person}{Benoit Baudry}, {and} \bibinfo{person}{Gurvan Le~Guernic}.} \bibinfo{year}{2016}\natexlab{}.
\newblock \showarticletitle{Reverse-engineering reusable language modules from legacy domain-specific languages}. In \bibinfo{booktitle}{\emph{Software Reuse: Bridging with Social-Awareness: 15th International Conference, ICSR 2016, Limassol, Cyprus, June 5-7, 2016, Proceedings 15}}. Springer, \bibinfo{pages}{368--383}.
\newblock


\bibitem[Mernik et~al\mbox{.}(2005)]%
        {10.1145/1118890.1118892}
\bibfield{author}{\bibinfo{person}{Marjan Mernik}, \bibinfo{person}{Jan Heering}, {and} \bibinfo{person}{Anthony~M. Sloane}.} \bibinfo{year}{2005}\natexlab{}.
\newblock \showarticletitle{When and how to develop domain-specific languages}.
\newblock \bibinfo{journal}{\emph{ACM Comput. Surv.}} \bibinfo{volume}{37}, \bibinfo{number}{4} (\bibinfo{date}{Dec.} \bibinfo{year}{2005}), \bibinfo{pages}{316–344}.
\newblock
\showISSN{0360-0300}
\urldef\tempurl%
\url{https://doi.org/10.1145/1118890.1118892}
\showDOI{\tempurl}


\bibitem[Meyerson et~al\mbox{.}(2023)]%
        {meyerson2023language}
\bibfield{author}{\bibinfo{person}{Elliot Meyerson}, \bibinfo{person}{Mark~J Nelson}, \bibinfo{person}{Herbie Bradley}, \bibinfo{person}{Adam Gaier}, \bibinfo{person}{Arash Moradi}, \bibinfo{person}{Amy~K Hoover}, {and} \bibinfo{person}{Joel Lehman}.} \bibinfo{year}{2023}\natexlab{}.
\newblock \showarticletitle{Language model crossover: Variation through few-shot prompting}.
\newblock \bibinfo{journal}{\emph{arXiv preprint arXiv:2302.12170}} (\bibinfo{year}{2023}).
\newblock


\bibitem[Saha and Narula(2011)]%
        {Saha2011GraminAS}
\bibfield{author}{\bibinfo{person}{Diptikalyan Saha} {and} \bibinfo{person}{Vishal Narula}.} \bibinfo{year}{2011}\natexlab{}.
\newblock \showarticletitle{Gramin: a system for incremental learning of programming language grammars}. In \bibinfo{booktitle}{\emph{International Symposium on Electronic Commerce}}.
\newblock
\urldef\tempurl%
\url{https://api.semanticscholar.org/CorpusID:1743193}
\showURL{%
\tempurl}


\bibitem[Suzuki et~al\mbox{.}(2023)]%
        {suzuki2023constructing}
\bibfield{author}{\bibinfo{person}{Masahiro Suzuki}, \bibinfo{person}{Hiroki Sakaji}, \bibinfo{person}{Masanori Hirano}, {and} \bibinfo{person}{Kiyoshi Izumi}.} \bibinfo{year}{2023}\natexlab{}.
\newblock \showarticletitle{Constructing and analyzing domain-specific language model for financial text mining}.
\newblock \bibinfo{journal}{\emph{Information Processing \& Management}} \bibinfo{volume}{60}, \bibinfo{number}{2} (\bibinfo{year}{2023}), \bibinfo{pages}{103194}.
\newblock


\bibitem[VIERU et~al\mbox{.}(2024)]%
        {vieru2024domain}
\bibfield{author}{\bibinfo{person}{Mihai VIERU}, \bibinfo{person}{Vlad POLISCIUC}, \bibinfo{person}{Daniela GLIGA}, \bibinfo{person}{Ecaterina GREBENNICOVA}, {and} \bibinfo{person}{Anastasia ZAGORODNIUC}.} \bibinfo{year}{2024}\natexlab{}.
\newblock \showarticletitle{Domain-specific language for analyzing medical results}.
\newblock  (\bibinfo{year}{2024}).
\newblock


\bibitem[Waite and Goos(2012)]%
        {waite2012compiler}
\bibfield{author}{\bibinfo{person}{William~M Waite} {and} \bibinfo{person}{Gerhard Goos}.} \bibinfo{year}{2012}\natexlab{}.
\newblock \bibinfo{booktitle}{\emph{Compiler construction}}.
\newblock \bibinfo{publisher}{Springer Science \& Business Media}.
\newblock


\bibitem[Ye et~al\mbox{.}(2023)]%
        {ye2023comprehensive}
\bibfield{author}{\bibinfo{person}{Junjie Ye}, \bibinfo{person}{Xuanting Chen}, \bibinfo{person}{Nuo Xu}, \bibinfo{person}{Can Zu}, \bibinfo{person}{Zekai Shao}, \bibinfo{person}{Shichun Liu}, \bibinfo{person}{Yuhan Cui}, \bibinfo{person}{Zeyang Zhou}, \bibinfo{person}{Chao Gong}, \bibinfo{person}{Yang Shen}, {et~al\mbox{.}}} \bibinfo{year}{2023}\natexlab{}.
\newblock \showarticletitle{A comprehensive capability analysis of gpt-3 and gpt-3.5 series models}.
\newblock \bibinfo{journal}{\emph{arXiv preprint arXiv:2303.10420}} (\bibinfo{year}{2023}).
\newblock


\bibitem[Ye and Durrett(2022)]%
        {ye2022unreliability}
\bibfield{author}{\bibinfo{person}{Xi Ye} {and} \bibinfo{person}{Greg Durrett}.} \bibinfo{year}{2022}\natexlab{}.
\newblock \showarticletitle{The unreliability of explanations in few-shot prompting for textual reasoning}.
\newblock \bibinfo{journal}{\emph{Advances in neural information processing systems}}  \bibinfo{volume}{35} (\bibinfo{year}{2022}), \bibinfo{pages}{30378--30392}.
\newblock


\end{thebibliography}

\appendix

\end{document}